# A comment on: the violations of locality and free choice are equivalent resources in Bell experiments.

Marian Kupczynski

Département de l'Informatique, Université du Québec en Outaouais , Case postale 1250, succursale Hull, Gatineau. QC, Canada J8X 3X7

**\* Correspondence:**
marian.kupczynski@uqo.ca

In {1 ) authors study a causal model $\{P_{ab|xy}\}_{xy}$ in which hidden variables $\lambda \in \Lambda$, distributed according to some $P_\lambda$, might have influenced Alice's and Bob's outcomes .

$$P_{ab|xy} = \sum_\lambda P_{ab|xy\lambda} P_{\lambda|xy} \qquad [1]$$

Local realistic model (LRHVM) is obtained by adding two assumptions:

- locality: $\qquad P_{ab|xy\lambda} = P_{a|x\lambda} P_{b|y\lambda} \qquad [2]$
- freedom of choice: $\qquad P_{\lambda|xy} = P_\lambda \qquad [3]$

Using LRHVM one derives CHSH inequalities, which are violated (2). The authors investigate *how often can a given assumptions i.e., locality or free choice, be retained, while safeguarding the other assumption, in order to fully reproduce some given experimental statistics* within a model [1-3]. They determine the maximal fraction $\mu_L$ of trials in which the locality may be retained and the maximal fraction $\mu_F$ of trials in which the assumption [3] may remain valid.

It turns out that $\mu_L = \mu_F$ and that they tend to 0 with growing number of settings. It is a troubling result because *locality* and *freedom of choice* should not be violated (2-4). Moreover, it is difficult to reject *realism* <u>as it is defined in (1)</u>: *physical objects and their properties exist, whether they are observed or not*

Fortunately the violation of Bell-type inequalities allows rejecting realism/CFD understood in much more restricted sense: *any system has pre-existing <u>values</u> for all possible measurements of the system* (5). Peres nicely concluded that: *unperformed experiments have no results* (6).

CFD implies the existence of a *counterfactual joint probability distribution* of all 4 random variables describing experimental outcomes and CHSH is a *noncontextuality inequality* for a 4-cyclic Bell-scenario (7). Fine (8) demonstrated that CHSH inequalities are the sufficient conditions for the existence of such probability distribution. In LRHVM the outcomes are determined by variables λ describing "twin-photon pairs" [3,4] and CHSH may not derived, if the assumption [3] is violated. Thus the assumption [3] should be called *CFD* or *noncontextuality* and not *freedom of choice*.

Bohr insisted that measuring instruments play an active role in creation of experimental outcomes. Therefore in contextual causal models (9, 10) a choice of a setting (x, y) is not a choice of two labels but a "choice" of $\Lambda_x \cup \Lambda_y$, where $\Lambda_x$ and $\Lambda_y$ are respective sets describing microstates of Alice's and Bob's PBS-detector modules at the moment of the interaction with incoming photonic beams.

In (9) we give a detailed proof, that the assumption [3], <u>contrary to what is often claimed</u>, should not be called *experimenters' freedom of choice*. Misunderstanding is rooted in incorrect interpretation of Bayes Theorem. Conditional probabilities do not allow, in general, any causal interpretation. For example, if an event A="xy"= $\Lambda_x \cup \Lambda_y$ and an event E={$\lambda_x$, $\lambda_y$}, then

$$P(E \cap A) = P(E) = P(E|A)P(A)\,;\; P(A|E) = P(E \cap A)/P(E) = 1 \qquad [4]$$

Experimenters may choose any experimental pair of settings (x, y), as they wish (2-4). The occurrence of a "hidden event E" tells us <u>only</u> that the pair (x, y) was chosen (9). In the notation of (1): $P_{xy|\lambda} = 1 \neq P_{xy}$. We have the *causal independence* without the *stochastic independence* $P_{xy\lambda} = P_{xy}P_{\lambda}$ [3].

Therefore the resource in Bell experiments is *contextuality* and not the violations of *locality* and/or *freedom of choice*. It is definitely the less mind boggling conclusion.